\documentclass[fleqn,10pt]{wlscirep}
\usepackage[utf8]{inputenc}
\usepackage[T1]{fontenc}
\title{Bound states  in the continuum (BIC) protected by self-sustained potential barriers in a flat band system}

\author[1*]{Yi-Cai Zhang}

\affil[1]{School of Physics and Materials Science, Guangzhou University, Guangzhou 510006, People’s
Republic of China}

\affil[*]{zhangyicai123456@163.com}

\begin{abstract}
 \textbf{In this work, we investigate the bound states in the continuum (BIC) of  a one-dimensional spin-1 flat band system.
It is found that, when the potential is sufficiently strong, there exists an effective attractive potential well surrounded by infinitely high self-sustained barriers.
Consequently, there exist some BIC in the effective potential well.
These bound states  are protected by the infinitely high potential barriers, which could not decay into the continuum.}
Taking a long-ranged  Coulomb potential and a short-ranged exponential potential as two examples, the bound state energies are obtained.
For a Coulomb potential, there exists a series of critical potential strengths, near which the bound state energy can go to infinity.
For a sufficiently strong exponential potential, there exists two different bound states with a same number of wave function nodes.
The existence of BIC protected by the self-sustained potential barriers is quite a universal phenomenon in the flat band system under a strong potential. A necessary condition for the existence of BIC is that the maximum value of potential is larger than two times band gap.
\end{abstract}
\begin{document}

\flushbottom
\maketitle
%
%
\thispagestyle{empty}


\section*{Introduction}

For usual potential wells, the bound states usually appear outside the continuous spectrum.
However, Neumann and Wigner constructed a bound state which is embedded in the continuous spectrum of scattering states \cite{Neumann1929} (the so-called bound states in the continuum (BIC)\cite{Stillinger1975}). Later, it is found that BIC can also appear due to the interferences of two resonances, where one of the resonance widths becomes zero with the variations of a continuous parameter \cite{Fano1961,Friedrich1985}.
BIC can also appear in the optical waveguides \cite{Bendix2009,Bulgakov2008,Longhi2010,Vicencio2015,Fong2017}, and condensed matter physics system \cite{zhangjiangmin,Molina2012,Gorbatsevich2017,Xiao2018,Takeichi2019}. In addition,  BIC can also exist in bottomless potentials \cite{Cho2008,Ahmed2019}.

A lot of  novel physics, for example, existences of localized flat band states \cite{Sutherland1986,Vidal1998,Mukherjee}, the ferro-magnetism transition \cite{Mielke1999,Zhang2010,Raoux2014}, super-Klein tunneling \cite{Shen2010,Urban2011,Fang2016,Ocampo2017}, preformed pairs \cite{Tovmasyan2018}, strange metal \cite{Volovik2019}, high $T_c$ superconductivity/superfluidity \cite{Peotta2015,Hazra2019,Cao2018,Wuyurong2021,Kopnin2011,Julku2020,Iglovikov2014,Julku2016,Liang2017,Iskin2019,Wu2021}, ect., can  appear in a flat band system.
Due to the existence of infinitely large density of states in a spin-1 flat band system, a short-ranged potential, e.g., square well potential, can result in infinite bound states, even a hydrogen atom-like  energy spectrum, i.e., $E_n\propto1/n^2,n=1,2,3,...$ \cite{Zhangyicai2021}.

Furthermore, it is found that the existences of  bound states also depend on the types of potentials.
 For example, a long-ranged Coulomb potential of type I (with three same diagonal elements in usual basis),
an arbitrary weak Coulomb potential can destroy completely the flat band \cite{Gorbar2019,Pottelberge2020}.
In two-dimensional spin-1 systems, a strong Coulomb potential can result in a wave function collapse near the the origin \cite{Gorbar2019,Han2019}. For one-dimensional case, an arbitrarily weak Coulomb potential also  causes the wave function collapse  \cite{Zhangyicai20212}.
In addition, for a potential of type II, which has a unique non-vanishing potential matrix element in  basis $|2\rangle$ \cite{Zhangyicai2021}, a long-ranged Coulomb potential can cause  a $1/n$ energy spectrum.
For a Coulomb potential of type III, which has a unique non-vanishing potential matrix element in  basis $|3\rangle$ \cite{Zhangyicai20213}, there are also infinite bound states which are generated from the flat band. Near the the flat band, the energy is inversely proportional to the natural number, i.e., $E\propto1/n$.
Differently from the ordinary one-dimensional bound state energy which is parabolic function of potential strength,  the bound state energy is linearly dependent on the potential strength as the strength goes to zero. For a given quantum number $n$, the bound state energy grows up with the increasing of potential strength $\alpha$.  There is  a critical potential strength $\alpha_{cr}$ at which  the bound state energy reaches the threshold  of continuous spectrum.  After crossing the threshold, these bound state may still exist and they would form the bound states in a continuous spectrum (BIC)  \cite{Zhangyicai20213}.

In this work, we propose a new mechanism of the existence of  BIC in a spin-1 flat band system with a  strong potential of type III.
 To be specific, we would extend the above investigations  (see Ref.\cite{Zhangyicai20213}) where the energy is in the continuum.
It is found that for sufficiently strong potential, there exist an effective potential well which are surrounded by infinitely high potential barriers. Within the potential well, there may exist some bound states which are embedded in the continuous spectrum, i.e., BIC. The infinitely high barriers protect the BIC from decaying into the continuous spectrum.
Taking a long-ranged Coulomb potential and a shorted-ranged exponential potential as two examples, we get the bound state (BIC) energies.
Our results show that the existence of BIC is quite a universal phenomenon for  a strong potential of type III in the spin-1 flat band system .


\subsection*{Results}

\subsection{The model Hamiltonian  with a flat band}
In this work, we consider a spin-1 Dirac-type Hamiltonian \cite{Zhangyicai2021} in one dimension, i.e.,
 \begin{align}
&H=H_0+V_p(x)\notag\\
&H_0=-iv_F\hbar S_x\partial_x+m S_z,
    	\label{hamiltonian}
\end{align}
where $V_p(x)$ is potential energy, $H_0$ is the free-particle Hamiltonian, $v_F>0$ is Fermi velocity, and $m>0$ is energy gap parameter. $S_x$ and $S_z$ are spin operators for spin-1 particles \cite{Zhang2013}, i.e.,
\begin{align}
&S_x=\left[\begin{array}{ccc}
0 &\frac{1}{\sqrt{2}}  & 0\\
\frac{1}{\sqrt{2}}& 0 &\frac{1}{\sqrt{2}}\\
0 &\frac{1}{\sqrt{2}} & 0
  \end{array}\right];&S_z=\left[\begin{array}{ccc}
1 &0  & 0\\
0&0& 0\\
0 &0 & -1
  \end{array}\right],
\end{align}
in usual spin basis $|i\rangle$ with $i=1,2,3$. In the whole manuscript, we use the units of $v_F=\hbar=1$.
 When $V_p(x)=0$, the free particle Hamiltonian $H_0$ has three energy bands. One of them is flat band with eigen-energy $E_{k,0}\equiv0$, and the other two of them are dispersion bands \cite{Zhangyicai2021}.  Among the three band, there are two band gaps, whose sizes are given by parameter $m$ (see Fig.1).
 Any possible BIC only exist in the two continuous spectrum, i.e., outside the gaps .

\begin{figure}
\begin{center}
\includegraphics[width=1.0\columnwidth]{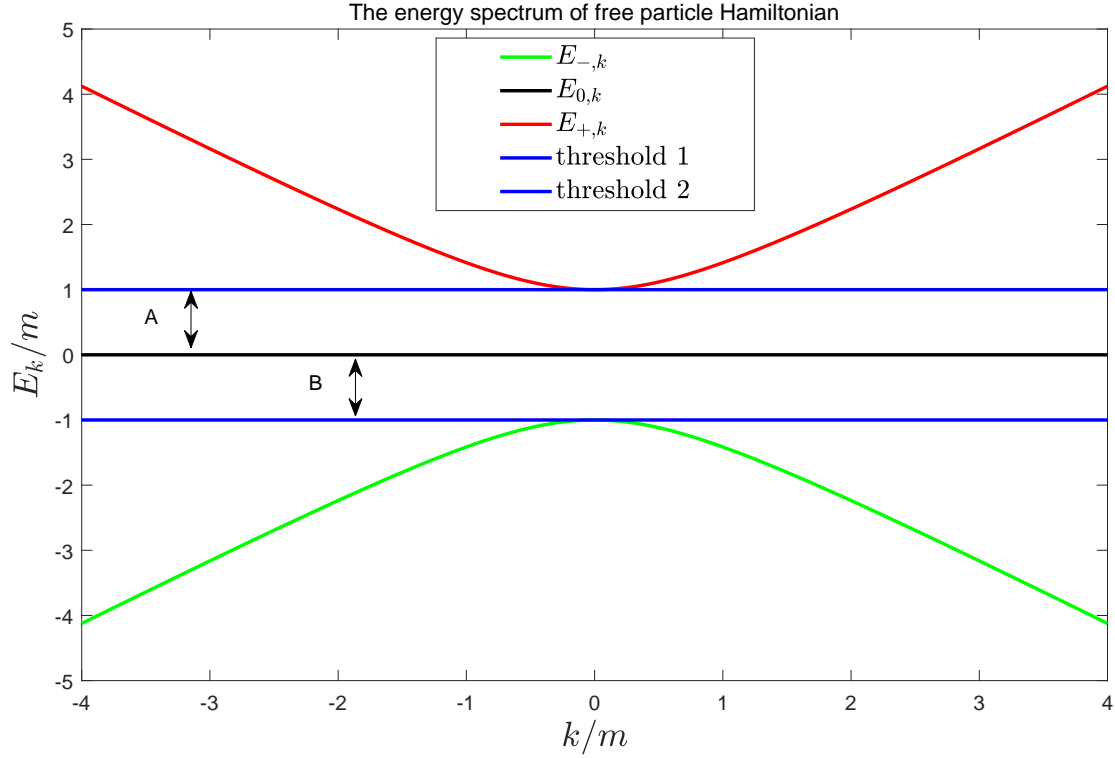}
\end{center}
\caption{ The energy spectrum of free particle Hamiltonian.
  The possible conventional bound states exist in the gaps \textbf{A} and \textbf{B}. Any possible bound states in the continuum (BIC) only exist outside these gaps.  }
\end{figure}

\subsection{bound states in a potential of type III}
In the following manuscript, we assume the potential energy $V_p$ has following form in usual basis $|i=1,2,3\rangle$, namely,
\begin{align}\label{3}
&V_p(x)=V_{11}(x)\bigotimes|1\rangle\langle1|
=\left[\begin{array}{ccc}
V_{11}(x) &0  & 0\\
0&0& 0\\
0 &0 & 0
  \end{array}\right].
\end{align}
In the whole manuscript, we would refer such a kind of potential as potential of type III \cite{Zhangyicai20213}. \textbf{ Such a spin-dependent
potential is a bit similar to the magnetic impurity potential in Kondo model. The conventional bound states for potentials of type I and II have also been investigated in our previous works \cite{Zhangyicai2021,Zhangyicai20212}.}
Adopting a similar procedure as Ref. \cite{Zhangyicai20213}, the spin-1 Dirac  equation
\begin{align}\label{4}
&-i\partial_x\psi_2(x)/\sqrt{2}=[E-m-V_{11}(x)]\psi_1(x),\notag\\
&-i\partial_x[\psi_1(x)+\psi_3(x)]/\sqrt{2}=E\psi_2(x),\notag\\
&-i\partial_x\psi_2(x)/\sqrt{2}=[E+m]\psi_3(x).
\end{align}
can be transformed into an effective Schr\"{o}dinger equation (a second-order differential equation), i.e.,
\begin{align}\label{eff}
-\partial_{x}^{2}\psi(x)+\tilde{V}(x)\psi(x)=\tilde{E}\psi(x).
\end{align}
where the  auxiliary wave function
\begin{align}\label{6}
\psi(x)\equiv\frac{E-V_{11}(x)/2}{E+m}\psi_{1}(x).
\end{align}
The  effective total energy $\tilde{E}$ and effective potential $\tilde{V}$ are
\begin{align}\label{7}
&\tilde{E}=E^2-m^2>0, \ for\  bound  \ states\ in \ Continuum,\notag\\
&\tilde{V}(x)=\frac{V_{11}(x)}{2}\frac{(m+E)^2}{E-V_{11}(x)/2}.
\end{align}
In the following, we would solve the effective Schr\"{o}dinger equation Eq.(\ref{eff}) to get the bound state (BIC) energies.

\subsubsection{a long-ranged Coulomb potential}
In this subsection, we assume the $V_{11}$ is  a Coulomb potential, i.e.,
\begin{align}\label{8}
V_{11}(x)=\frac{\alpha}{|x|},
\end{align}
where $\alpha$ is the potential strength.
The effective potential $\tilde{V}$ is
\begin{align}\label{9}
\tilde{V}(x)=\frac{V_{11}(x)}{2}\frac{(m+E)^2}{E-V_{11}(x)/2}=\frac{A}{|x|-x_0}.
\end{align}
In the above equation, we introduce parameter $A\equiv\frac{\alpha (m+E)^2}{2E}$ and $x_0\equiv\frac{\alpha}{2E}$. It is shown that the effective potential $\tilde{V}$ is a shifted Coulomb potential  \cite{Downing2014} with an effective potential strength $A$, which depends on energy $E$.
The Eq.(\ref{eff}) becomes
\begin{align}\label{Coul}
\partial_{x}^{2}\psi(x)+[\tilde{E}-\frac{A}{|x|-\frac{\alpha}{2E}}]\psi(x)=0.
\end{align}

\begin{figure}
\begin{center}
\includegraphics[width=1.0\columnwidth]{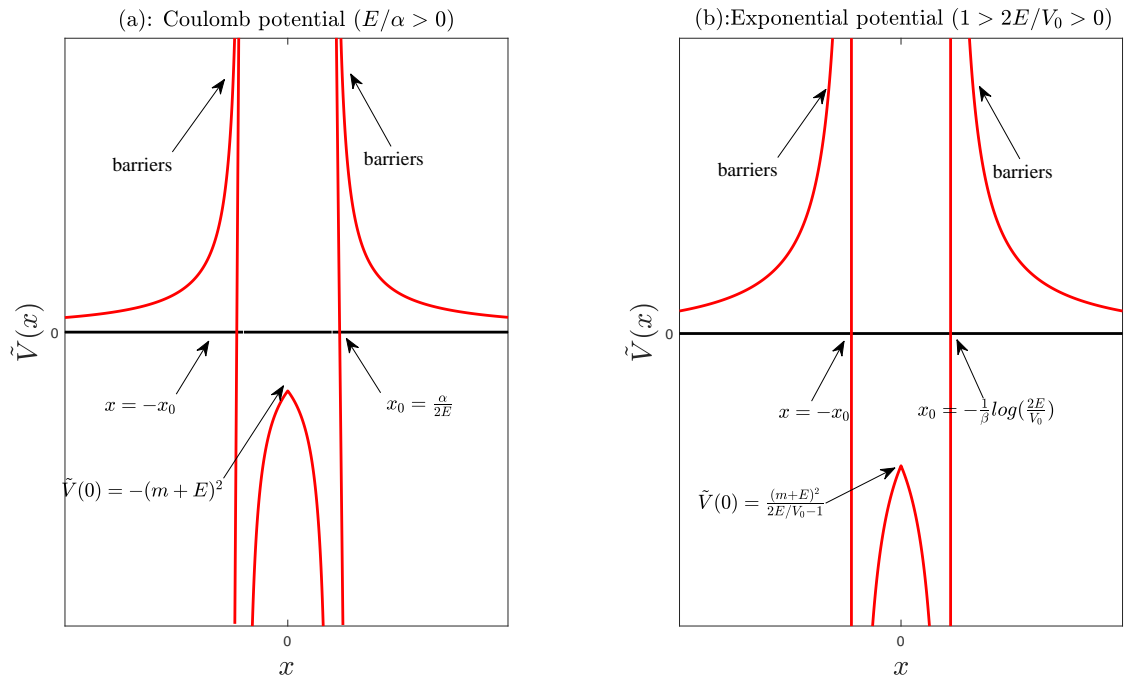}
\end{center}
\caption{ The effective potential wells surrounded by two infinitely high barriers (the red solid lines). (a): The effective potential for Coulomb potential ($\alpha/E>0$). The value of effective potential at $x=0$, i.e., $\tilde{V}(0)=-(m+E)^2$. (b): The effective potential for exponential potential.  The value of effective potential at $x=0$, i.e., $\tilde{V}(0)=\frac{(m+E)^2}{\frac{2E}{V_0}-1}$}
\end{figure}

Depending on the sign of $\alpha/E$, there exists two kinds of effective potentials $\tilde{V}$ \cite{Zhangyicai20213}.
 When $\alpha/E<0$,   the effective potential $\tilde{V}$ has a lowest point at $x=0$. The bound state energies are in the the gaps, $0<E<m$. Hence there is no BIC for $\alpha/E<0$.

When $\alpha/E>0$ ($x_0>0$), the effective potential $\tilde{V}$ is negative in the interval $(-x_0,x_0)$ (a potential well), and positive in intervals $(-\infty,-x_0)$ and $(x_0,\infty)$ (see Fig.2). There are two infinitely high potential barriers near two ends  $x=\pm x_0$ of the interval $(-x_0,x_0)$.
We see that the potential well width $w\equiv 2x_0=\alpha/E$, and the position of potential barriers $x=\pm x_0$ depend sensitively on the bound state energy $E$. So in the whole work, we call the potential barriers as self-sustained potential barriers.
 In addition, it is found that the bound state energy can be larger than zero for  $\alpha>0$ or  smaller than zero for  $\alpha<0$ \cite{Zhangyicai20213}.
 In the following, we mainly focus on the case of $\alpha/E>0$, where the bound states in the continuous spectrum (BIC) may appear.

When $x>0$ and $|E|>m$ (for BIC), the equation Eq.(\ref{Coul}) can be solved with some confluent hypergeometric functions. Its general solution is
\begin{align}\label{11}
&\psi(x)=(x-x_0)e^{-i\sqrt{\tilde{E}}(x-x_0)}\{c_{1}\times {}_1F_1[a,b,2i\sqrt{\tilde{E}}(x-x_0)]+c_{2}\times U[a,b,2i\sqrt{\tilde{E}}(x-x_0)]\},
\end{align}
where ${}_1F_1[a,b,z]=\sum_{k=0}^{\infty}\frac{(a)_kz^k}{k!(b)_k}$ is confluent hypergeometric function \cite{Abramowitz},  $(a)_k=a\times (a+1)\times(a+2)\times...\times(a+k-1)$, and $c_1(c_2)$ are two arbitrary constants. $a=1+\frac{A}{2i\sqrt{\tilde{E}}}$, $b=2$. $U[a, b, z]$ is
a second linearly independent solution to the confluent
hypergeometric equation (Tricomi function \cite{Wang1989}).
When $z\rightarrow 0$, the two confluent hypergeometric functions behave as
\begin{align}\label{12}
&{}_1F_1[a,b,z]\simeq1,\notag\\
&U[a,b,z]= \frac{\Gamma(b-1)}{\Gamma(a)}z^{1-b}+O(|logz|), \quad (b=2).
\end{align}

When $\alpha/E>0$, due to the existence of infinitely high  potential barriers, the bound state in the continuum  only exist in the interval $(-x_0,x_0)$ (see Fig.2). Outside the effective potential well, the wave function vanishes (see Fig.5). At the two ends of the interval, the zero boundary conditions should be satisfied, i.e.,
\begin{align}\label{13}
&\psi(\pm x_0)=0.
\end{align}
Taking  Eq.(\ref{12}) into account, $U[a,b,2i\sqrt{\tilde{E}}(x-x_0)]$ should be discarded. So the wave function is
\begin{align}\label{14}
&\psi(x)=(x-x_0)e^{-i\sqrt{\tilde{E}}(x-x_0)}{}_1F_1[a,b,2i\sqrt{\tilde{E}}(x-x_0)].
\end{align}

Due to $\tilde{V}(-x)=\tilde{V}(x)$, the wave functions can be classified by parities.
For odd parity states, the bound state energy equation is
\begin{align}\label{15}
{}_1F_1[1+\frac{\alpha(E+m)^2}{4iE\sqrt{E^2-m^2}},2,-\frac{i\alpha\sqrt{E^2-m^2}}{E}]=0.
\end{align}
For even parity states, the bound state energy equation is
\begin{align}\label{16}
&[-4E(2E+i\alpha\sqrt{E^2-m^2})]\times{}_1F_1[1+\frac{\alpha(E+m)^2}{4iE\sqrt{E^2-m^2}},2,-\frac{i\alpha\sqrt{E^2-m^2}}{E}]\notag\\
&+\alpha[4i E\sqrt{E^2-m^2}+\alpha(m+E)^2]\times{}_1F_1[2+\frac{\alpha(E+m)^2}{4iE\sqrt{E^2-m^2}},3,-\frac{i\alpha\sqrt{E^2-m^2}}{E}]=0.
\end{align}
The results are reported in Fig.(3).

\begin{figure}
\begin{center}
\includegraphics[width=1.0\columnwidth]{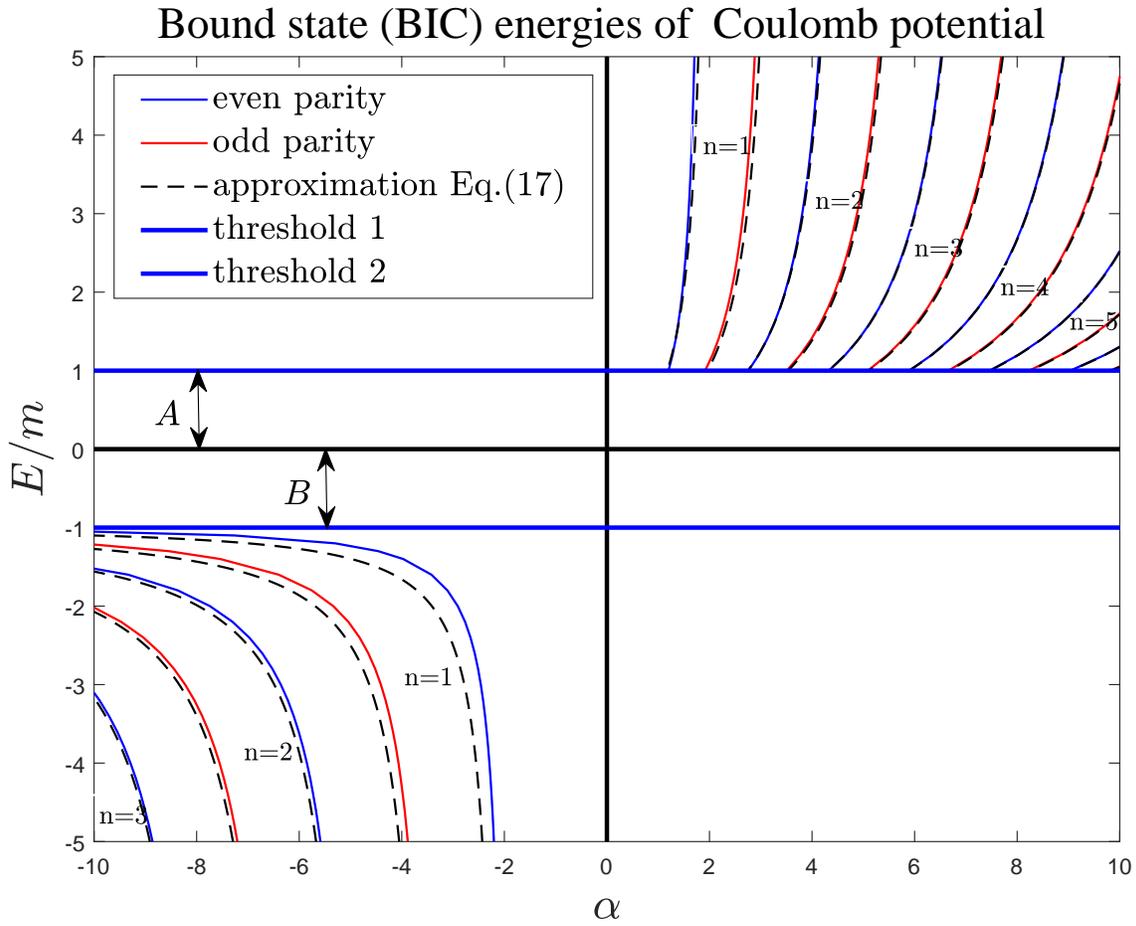}
\end{center}
\caption{ The bound state (BIC) energy of Coulomb potential in the case of $\alpha/E>0$. The solid lines are the the exact results of Eqs.(\ref{15}) and (\ref{16}). The black dashed lines are given by the quasi-classical approximation formula Eq.(\ref{17}). \textbf{In the gaps A and B, there also exist infinite  (conventional) bound states whose energy satisfy $|E|<m$  (see Fig. 4 of Ref. [45]}.)    }
\end{figure}

When $E^2>m^2$ for BIC, with quasi-classical approximation method \cite{Landau},  the eigen-energy  is given by
\begin{align}\label{17}
\frac{\alpha|m+E|}{2E}[\sqrt{\frac{2E}{m+E}}+\sqrt{\frac{m+E}{E-m}}arcsinh(\sqrt{\frac{E-m}{E+m}})]=(n+\Delta)\pi,
\end{align}
where $n=1,2,3,....$, and
\begin{align}\label{18}
&\Delta=+\frac{1}{4},\ for\ odd \ parity \ states,\notag\\
&\Delta=-\frac{1}{4},\ for\ even \ parity \ states.
\end{align}
From the Fig. 3, we see that the the bound state (BIC) energy can be well described by the quasi-classical approximation Eq.(\ref{17}).

For $\alpha>0$, there exists a critical potential strength \cite{Zhangyicai20213}, which is determined by
\begin{align}\label{19}
&\alpha_{cr1}\simeq\frac{(1/4+n)\pi}{2},\ for\ odd \ parity \ states,\notag\\
&\alpha_{cr1}\simeq\frac{(-1/4+n)\pi}{2},\ for\ even \ parity \ states.
\end{align}
 After crossing these critical values, the bound states still exist and they  form the bound states in continuum (BIC).
For a given $n$, there exist another critical value $\alpha_{cr2}$, near which the bound state energy goes to infinity, i.e., $E\rightarrow\pm\infty$.
 When $E\rightarrow\pm\infty$, the value is can be determined approximately with Eq.(\ref{17}), i.e.,
\begin{align}\label{20}
&\alpha_{cr2}\simeq \pm\frac{2(n+\Delta)\pi}{\sqrt{2}+arcsinh(1)}\simeq\pm0.6534(n+\Delta)\pi,
\end{align}
where sign $+(-)$ for $E>(<)0$.
For $\alpha<0$, when $E<-m$ and $|\alpha|\gg1$, for a given $n$, the energy approach the threshold of lower band, i.e.,  $E\rightarrow-m$.

The existence of the BIC can be understood qualitatively  as follows.
The width of  effective potential well is $w\equiv2x_0=\alpha/E$.  Based on the bound state energy formulas of infinitely deep square well potential,  the bound state energy equation can be approximately written as
 \begin{align}\label{21}
kw= n\pi,
\end{align}
where wave vector can be approximated by
 \begin{align}\label{22}
k\sim \sqrt{\tilde{E}-\tilde{V}(0)}=\sqrt{2E(m+E)}.
\end{align}
So we find that the potential strength
\begin{align}\label{23}
\alpha= \frac{n \pi E}{\sqrt{2E(m+E)}},
\end{align}
where $E^2>m^2$ for BIC.
Only when
\begin{align}\label{24}
&\alpha>\alpha_{cr1}\simeq \pi/2, \ for  \ E>m,
\end{align}
 BIC begin to appear. When $E\rightarrow\pm\infty$, the critical value
\begin{align}\label{25}
&\alpha_{cr2}\simeq\pm n\pi/\sqrt{2}\simeq\pm0.7071n\pi.
\end{align}
A more accurate formula Eq.(\ref{20}) gives $\alpha_{cr2}\simeq\pm0.6534(n+\Delta)\pi$.

\subsubsection{short-ranged exponential potential}

We should remark that the above mechanism of existence of BIC is quite a universal phenomenon for a sufficiently strong potential.
 As long as the self-sustained potential barriers can form, the BIC may exist.  A critical condition for the existence of the infinitely high potential barriers is that the denominator of the effective potential has zero point [see Eq.(\ref{7})], i.e., the equation
 \begin{align}\label{26}
E-V_{11}(x)/2=0
\end{align}
has real solutions  $x$.
On the other hand, for BIC, the bound state energy should satisfy
 \begin{align}\label{27}
|E|>m.
\end{align}
 Taking the Eqs.(\ref{26}) and (\ref{27}) into account, we conclude that
 when the maximum value of potential is larger than two times gap parameter $m$, i.e.,
 \begin{align}\label{28}
|V_{11}(x)|_{max}>2m,
\end{align}
there may exist BIC.
We should emphasize that the above condition Eq.(\ref{28}) just is  a necessary condition for the existence of BIC.
So when the potential is sufficiently strong, there may exist BIC in the flat band system.

In this subsection, we assume the potential have an exponential function form, i.e.,
\begin{align}\label{29}
V_{11}(x)=V_0 e^{-\beta|x|},
\end{align}
where $V_0$ is the potential strength, $1/\beta$ describes decaying distance of exponential function.
 For the above sufficiently strong  exponential potential, it is found that there exist BIC.
The effective potential $\tilde{V}$ is
\begin{align}\label{30}
\tilde{V}=\frac{V_{11}(x)}{2}\frac{(m+E)^2}{E-V_{11}(x)/2}=\frac{(m+E)^2e^{-\beta|x|}}{\gamma-e^{-\beta|x|}},
\end{align}
where we introduce dimensionless parameter $\gamma\equiv \frac{2E}{V_0}$.
When $\gamma\equiv \frac{2E}{V_0}<1$, $\gamma-e^{-\beta|x|}=0$, i.e., $x=\pm x_0\equiv\mp\frac{1}{\beta}log(\gamma)$, then the denominator of effective potential is zero.
In the interval $(-x_0,x_0)$, the effective potential is negative, while outside the interval, the effective potential is positive. Near the two end of interval, i.e., $x=x_0$, there are also two infinitely high potential barriers (see Fig.2). So the bound states in continuum  may exist in the interval.

The Eq.(\ref{eff}) becomes
\begin{align}\label{31}
\partial_{x}^{2}\psi(x)+[\tilde{E}-\frac{\varepsilon e^{-\beta|x|}}{\gamma-e^{-\beta|x|}}]\psi(x)=0.
\end{align}
where $\tilde{E}=E^2-m^2$, $\varepsilon=(m+E)^2$.
The general solution of Eq.(\ref{31}) is
\begin{align}\label{32}
\psi(x)=c_{1}(\gamma e^{\beta x})^{-\frac{\sqrt{-(\tilde{E}+\varepsilon)}}{\beta}} {}_2F_1[A_-,B_-;C_-,\gamma e^{\beta x}]+c_{2}(\gamma e^{\beta x})^{\frac{\sqrt{-(\tilde{E}+\varepsilon)}}{\beta}}{}_2F_1[A_+,B_+;C_+,\gamma e^{\beta x}],
\end{align}
where ${}_2F_1[a,b;c,z]=\sum_{k=0}^{\infty}\frac{(a)_k(b)_kz^k}{k!(c)_k}$ is hypergeometric function,  $(a)_k=a\times (a+1)\times(a+2)\times...\times(a+k-1)$, and $c_1(c_2)$ are two arbitrary constants. $A_{\mp}=\mp\frac{\sqrt{-(\tilde{E}+\varepsilon)}}{\beta}-\frac{\sqrt{-\tilde{E}}}{\beta}$, $B_{\mp}=\mp\frac{\sqrt{-(\tilde{E}+\varepsilon)}}{\beta}+\frac{\sqrt{-\tilde{E}}}{\beta}$, $C_{\mp}=1\mp\frac{2\sqrt{-(\tilde{E}+\varepsilon)}}{\beta}$.
 At two ends of interval $(-x_0,x_0)$, the wave function should vanish, i.e.,
\begin{align}\label{33}
\psi(x=x_0)=c_{1}(\gamma e^{\beta x_0})^{-\frac{\sqrt{-(\tilde{E}+\varepsilon)}}{\beta}} \times{}_2F_1[A_-,B_-;C_-,\gamma e^{\beta x_0}]+c_{2}(\gamma e^{\beta x_0})^{\frac{\sqrt{-(\tilde{E}+\varepsilon)}}{\beta}}\times{}_2F_1[A_+,B_+;C_+,\gamma e^{\beta x_0}]=0.
\end{align}
\begin{figure}
\begin{center}
\includegraphics[width=1.0\columnwidth]{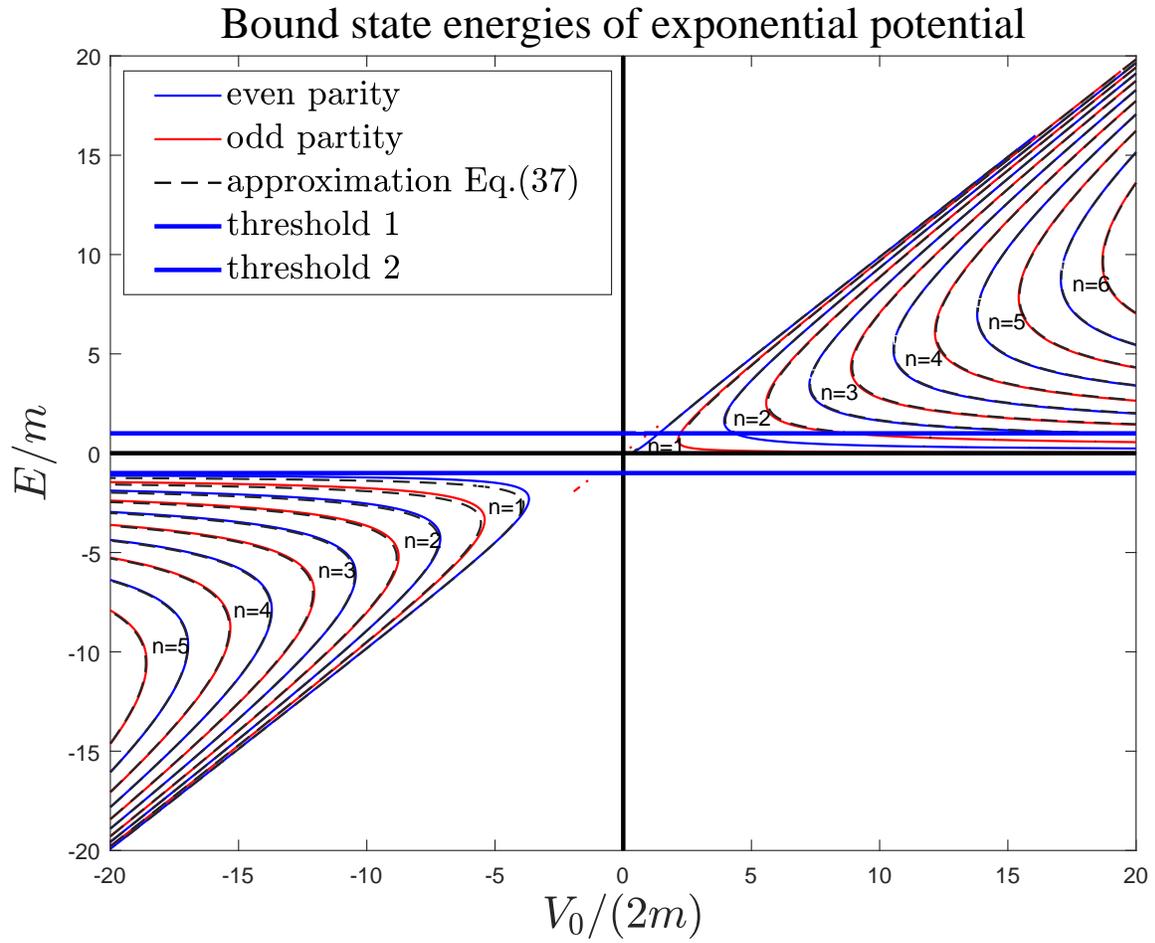}
\end{center}
\caption{ The bound state (BIC) energy of exponential potential. The solid lines are the the exact results of Eqs. (\ref{33}), (\ref{35}), and (\ref{36}) . The black dashed lines are given by the quasi-classical approximation formula Eq. (\ref{37}). Here we take $\beta=m$. \textbf{We note that there also exist some conventional bound states in the gaps.}  }
\end{figure}

Similarly, the bound state energy equation are given by
\begin{align}\label{34}
&\psi(x=0)=0,\ for\ odd \ parity \ states,\notag\\
&\psi'(x=0)=0,\ for\ even \ parity \ states.
\end{align}
To be specific, for odd parity states, the bound state energy equation is
\begin{align}\label{35}
c_{1}\gamma ^{-\frac{\sqrt{-(\tilde{E}+\varepsilon)}}{\beta}}\times {}_2F_1[A_-,B_-;C_-,\gamma ]+c_{2}\gamma ^{\frac{\sqrt{-(\tilde{E}+\varepsilon)}}{\beta}}\times{}_2F_1[A_+,B_+;C_+,\gamma ]=0.
\end{align}
For even parity states, the bound state energy equation is
\begin{align}\label{36}
&\frac{\gamma^{-\frac{\sqrt{-\tilde{E}-\varepsilon}}{\beta}}c_1}{-\beta+2\sqrt{-(\tilde{E}+\varepsilon)}}\{[2(\tilde{E}+\varepsilon)+\beta\sqrt{-\varepsilon-\tilde{E}}]\times{}_2F_1[A_-,B_-;C_-,\gamma]+\varepsilon\gamma \times{}_2F_1[1+A_-,1+B_-;1+C_-,\gamma]\}\notag\\
&-\frac{\gamma^{\frac{\sqrt{-\tilde{E}-\varepsilon}}{\beta}}c_2}{\beta+2\sqrt{-(\tilde{E}+\varepsilon)}}\{[2(\tilde{E}+\varepsilon)-\beta\sqrt{-\varepsilon-\tilde{E}}]\times{}_2F_1[A_+,B_+;C_+,\gamma]+\varepsilon\gamma\times{}_2 F_1[1+A_+,1+B_+;1+C_+,\gamma]\}=0
\end{align}
The results are reported in Fig.(4).

With quasi-classical approximation method,  the eigen-energy  is given by
\begin{align}\label{37}
&\frac{-2\sqrt{E^2-m^2}}{\beta}log[\sqrt{\frac{(V_0/2-E)(E-m)}{V_0(E+m)/2}}+\sqrt{\frac{mE+V_0 E-E^2}{V_0(E+m)/2}}]+\frac{\sqrt{2E(E+m)}}{\beta}log[\frac{1+\sqrt{(m+V_0-E)/(V_0-2E)}}{-1+\sqrt{(m+V_0-E)/(V_0-2E)}}]\notag\\
&=(n+\Delta)\pi,
\end{align}
where $n=1,2,3,....$, and
\begin{align}\label{38}
&\Delta=+\frac{1}{4},\ for\ odd \ parity \ states,\notag\\
&\Delta=-\frac{1}{4},\ for\ even \ parity \ states.
\end{align}

From Fig.4, it shows that for a given $V_0$  there exists two bound states with same quantum number $n$. \textbf{The two wave functions of $n=3$ with same number nodes are reported in panel (b) of Fig.5.}
  When potential is very strong, i.e., $V_0\rightarrow\pm\infty$, one of  bound state energy $E\rightarrow V_0/2$  (see Fig.4). The other bound state energy
\begin{align}
&E\rightarrow 0,\ for \ E>0\notag\\
&E\rightarrow -m,\ for \ E<0.
\end{align}

Similarly,  the above behaviors of bound state  energy can be explained as follows.
Based on the existence conditions of bound states of infinitely deep square well potential, when
 \begin{align}
kw= n\pi
\end{align}
where effective well width $w\equiv2x_0=-2\frac{1}{\beta}log(\gamma)$, the bound states would exist. The wave vector can be approximated by
 \begin{align}
k\sim \sqrt{\tilde{E}-\tilde{V}(0)}=\sqrt{\frac{2E(m+E)(E-m-V_0)}{2E-V_0}}
\end{align}
Therefore, the bound state energy equation can be approximately by
\begin{align}\label{42}
\frac{2}{\beta}\sqrt{\frac{2E(E+m)(E-m-V_0)}{2E-V_0}}log[\frac{V_0}{2E}]=n\pi
\end{align}
where $V_0/(2E)>1$ .
When the potential is very strong, e.g., $V_0\rightarrow\pm\infty$, for a given $n$,
the asymptotic behaviors of bound state energy can be obtained with Eq.(\ref{42}).
%
\begin{figure}
\begin{center}
\includegraphics[width=1.0\columnwidth]{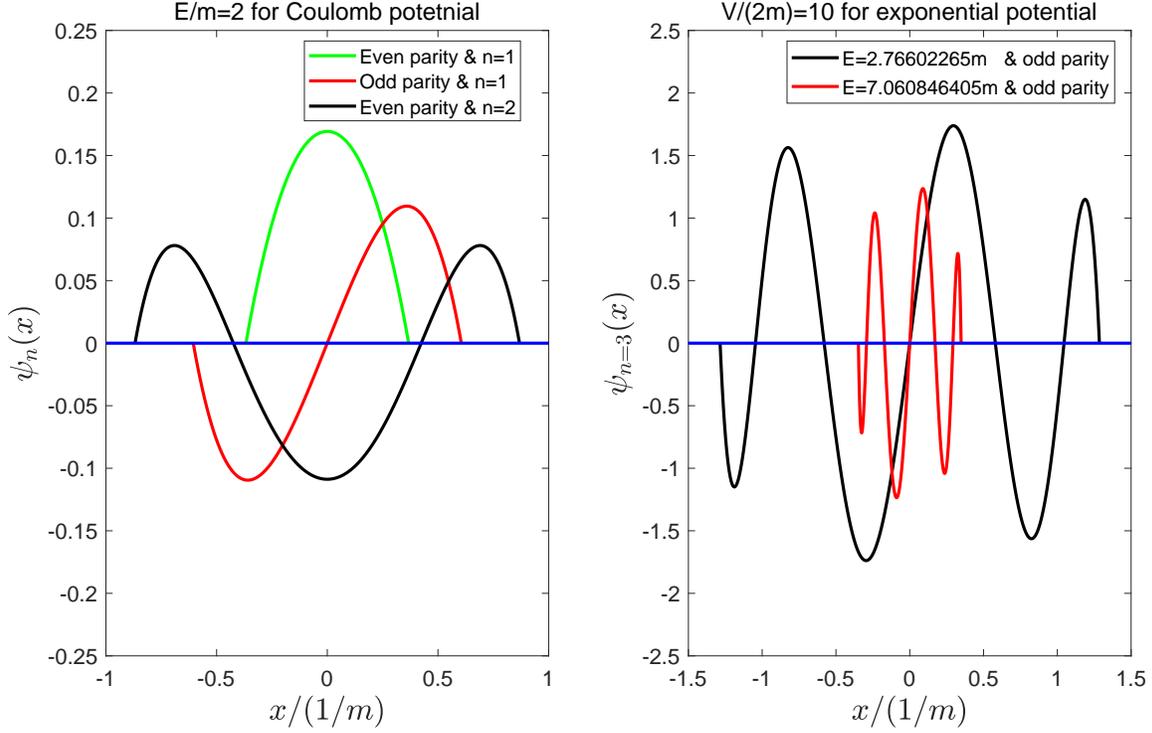}
\end{center}
\caption{ \textbf{The (un-normalized) wave functions for BIC.  (a): the
wave functions for Coulomb potential. Here we take the bound state energy $E/m=2$ for all the wave functions. (b): the two wave functions of $n=3$ with same number nodes for  exponential potential with potential strength $V/(2m)=10$. Here we take $\beta=m$. From Fig.5, we can see that outside the effective potential well which is given by ($-x_0<x<x_0$),  the wave functions vanish. } }
\end{figure}

\textbf{Finally, we note that for power-law decaying potential, i.e., $V_{11}(x)=\alpha/|x|^\delta$ with $\delta>0$, if  $\alpha>0$, only when $1<\delta<2$, the system can have  two bound states with same number wave function nodes. While for $\alpha<0$, only when $0<\delta<1$, the system has two bound states with same number nodes.}

\section*{summary}
In conclusion, we investigate the bound states in the continuum (BIC)  of  a one-dimensional spin-1 flat band system.
 It is found that BIC can exist for sufficiently strong potentials of type III. We get the bound state energies for a Coulomb potential and an exponential potential.
 For a Coulomb potential and a given quantum number $n$, when the potential strength reaches a critical value $\alpha_{cr1}$, the BIC begin to appear. When the potential strength reaches $\alpha_{cr2}$, the bound state energy goes to infinite.
 For exponential potential, there are two bound states with same number of wave function nodes.
 When the exponential potential is very strong, one of bound state energy approaches one half of  the potential strength.
 For repulsive potential (positive $V_0$),  the other bound state energy goes to zero.
For attractive case, the other bound state energy approaches to the threshold of lower continuous spectrum. In addition, it is found that  the bound state energies can be well described by the quasi-classical approximation.

 A necessary condition for existence of BIC is that the maximum value of of potential is larger than two times band gap. Our results shows that the existence of  BIC is quite a universal phenomenon for a strong potential of type III in the flat band system.

\bibliography{sample}



\section*{Acknowledgements (not compulsory)}
 Y.C. Z. are supported by the NSFC under Grants No.11874127 and startup grant from Guangzhou University.

\section*{Author contributions statement}
 Y.C.Z.  completed the paper.

\section*{Additional information}

Competing interests: The authors declare no competing interests





\end{document}